\documentclass[sigplan,screen]{acmart}

\renewcommand\footnotetextcopyrightpermission[1]{} 
\usepackage{awesomebox}
\usepackage{multirow}
\usepackage[verbatim]{lstfiracode}
\usepackage{mathastext}
\definecolor{halfgray}{HTML}{2B91AF}
\definecolor{j_frame}{HTML}{D3D4D5}
\definecolor{j_bg}{HTML}{FFFFFE}
\definecolor{j_red}{HTML}{699E08}
\definecolor{j_green}{HTML}{F9025A}
\definecolor{j_cyan}{HTML}{867B67}
\definecolor{j_purple}{RGB}{170, 34, 255}
\usepackage{listings}
\lstdefinelanguage[]{iJava}[]{java}{
    xleftmargin={0.75cm},
    sensitive=true,%
    commentstyle=\color{j_cyan}\ttfamily,
    stringstyle=\color{j_red}\ttfamily,
    keepspaces=true,
    showspaces=false,
    showstringspaces=false,
    rulecolor=\color{j_frame},
    frame=single,
    frameround={t}{t}{t}{t},
    framexleftmargin=6mm,
    numbers=left,
    numberstyle=\tiny\color{halfgray},
    backgroundcolor=\color{j_bg},
    style=FiraCodeStyle,
    basicstyle=\scriptsize\ttfamily,
    keywordstyle=\color{j_green}\ttfamily,
}
\begin{document}

\title{LLbezpeky: Leveraging Large Language Models for Vulnerability Detection}

\author{Noble Saji Mathews}
\orcid{0000-0003-2266-8848}
\affiliation{%
  \institution{University of Waterloo, Canada}
   \country{}
}
\email{noblesaji.mathews@uwaterloo.ca}

\author{Yelizaveta Brus}
\affiliation{%
  \institution{University of Waterloo, Canada}
   \country{}
}
 \email{ybrus@uwaterloo.ca}
 \author{Yousra Aafer}
\affiliation{%
  \institution{University of Waterloo, Canada}
   \country{}
}
 \email{yousra.aafer@uwaterloo.ca}
 \author{Meiyappan Nagappan}
\affiliation{%
  \institution{University of Waterloo, Canada}
   \country{}
}
 \email{mei.nagappan@uwaterloo.ca}
  \author{Shane McIntosh}
\affiliation{%
  \institution{University of Waterloo, Canada}
   \country{}
}
 \email{shane.mcintosh@uwaterloo.ca}
\acmConference[CS858]{}{Project Proposal}{LLbezpeky}
\begin{abstract}
Despite the continued research and progress in building secure systems, Android applications continue to be ridden with vulnerabilities, necessitating effective detection methods. Current strategies involving static and dynamic analysis tools come with limitations like overwhelming number of false positives and limited scope of analysis which make either difficult to adopt. Over the past years, machine learning based approaches have been extensively explored for vulnerability detection, but its real-world applicability is constrained by data requirements and feature engineering challenges. Large Language Models (LLMs), with their vast parameters, have shown tremendous potential in understanding semnatics in human as well as programming languages. We dive into the efficacy of LLMs for detecting vulnerabilities in the context of Android security. We focus on building an AI-driven workflow to assist developers in identifying and rectifying vulnerabilities. Our experiments show that LLMs outperform our expectations in finding issues within applications correctly flagging insecure apps in 91.67\% of cases in the Ghera benchmark. We use inferences from our experiments towards building a robust and actionable vulnerability detection system and demonstrate its effectiveness. Our experiments also shed light on how different various simple configurations can affect the True Positive (TP) and False Positive (FP) rates.
\end{abstract}
\maketitle

\section{INTRODUCTION}
Despite advancements in building secure systems and extensive research in the area, Android applications remain prone to a range of vulnerabilities and even vulnerability reintroduction for fixed issues \cite{gao2019understanding, almanee2021too}, creating a pressing demand for effective vulnerability detection methodologies. Current approaches to tackle this primarily revolve around static and dynamic analysis tools \cite{senanayake2023android}. However, they have their distinct limitations, such as vast false positives in the case of static analysis tools \cite{chao2020android} and an overwhelming amount of effort that goes into building these frameworks and adapting them to newer vulnerability types. 

In the past few years, there have also been numerous explorations into the use of machine learning to uncover vulnerabilities \cite{senanayake2023android}, however, the applicability of these remains limited in real-world settings due to the vast amount of training data required and an explicit focus on feature engineering to approximate the complexity of the systems being analyzed. With the advent of LLMs, huge billion parameter models have pushed the boundaries of what was thought achievable through an Artificially Intelligent system. These are believed to acquire ``embodied knowledge about syntax, semantics, and ontology inherent in human language'' \cite{manning2022human} and have also shown significant power when dealing with programming languages due to their relatively simpler underlying Grammar and Semantics \cite{hou2023large}. This brings a question to mind, How good are these Language Models at detecting vulnerabilities?. There have been several recent explorations into the use of these large language models and improving their efficacy for vulnerability detection in general, and these have shown promising results \cite{noever2023can}. 

Current work in the literature has looked at the performance of GPT-based models for the task of vulnerability detection in code \cite{cheshkov2023evaluation}, Cheskov et al. report it to be not very effective but they employ a very simplistic approach. Other recent work has shown that extended prompting and LLM-driven methods complemented by other techniques have yielded more accurate results than simple prompting for detecting CWEs present in code \cite{wang2023defecthunter, asare2023security}. There has also been a much more detailed exploration that looks into various aspects of using LLMs for software security with promising results \cite{zhang2023prompt}.

We explore this area further in the context of Android Security, attempting to build an AI-enabled workflow that would enable developers to detect and aide in remediating vulnerabilities or even aid developers in writing secure code by acting as a pair programmer. 
Do note that fine-tuning models or training our own LLM is out of scope of this research, we primarily seek to explore if and how the latent knowledge present in these LLMs can be used for Android analysis.

For this exploration into the use of new and emerging technologies it is critical we lay out a few fundamental questions, to begin with, and direct our efforts. We attempt to answer as many of the following as possible based on insights gained from this exploration:

\begin{itemize}
    \item \textbf{RQ1: Can LLMs detect Android vulnerabilities with basic prompting techniques?}
    \begin{itemize}
        \item If so how good are they at this compared to existing tools?
        \item What kind of vulnerabilities are LLMs good at uncovering? Can these systems detect newer vulnerabilities that are less well-known?
        \item Do we require fine-tuning of either the Model or the embeddings to achieve better results? 
    \end{itemize} 
    \item \textbf{RQ2: What kind of Input would such a system need to be able to understand and discover issues in the code?}
    \begin{itemize}
        \item How could we supply this additional context? What kind of Knowledge bases would aid LLMs in discovering complicated bugs? (API Usage, Permissions Involved, Framework Endpoints, Cross-language and dynamic components)
        \item Can existing solutions / static tools be used in tandem with these LLMs? or is it better to address only some classes of vulnerabilities with each type of approach?
    \end{itemize} 
\end{itemize}

To build out such a workflow there are 2 key elements we need to address (We give a brief overview of these in the following sections)

\begin{itemize}
    \item How do we get LLMs (particularly Instruct Trained Models) to do what we want? -> \textbf{Prompt Engineering}
    \item How do we supply the code and additional context to an LLM -> \textbf{Retrieval Augmented Generation}
\end{itemize}

\subsection{PROMPT ENGINEERING}
Prompt Engineering, a novel technique in artificial intelligence, is instrumental in enhancing the efficacy of language models for specific tasks. It involves intricate prompt construction that optimizes AI performance by eliciting improved responses \cite{white2023prompt}. One groundbreaking strategy within Prompt Engineering is the Chain-of-Thought Prompting introduced by Wei et al. (2022) \cite{wei2022chain}. This approach pushes the boundary of AI's reasoning by guiding the model through a sequence of prompts that enrich and build upon each other. It allows for more depth in AI reasoning, particularly when paired with few-shot prompting, proving useful for complex tasks that necessitate multiple stages of reasoning.


\subsection{RETRIEVAL-AUGMENTED GENERATION}
Retrieval Augmented Generation (RAG) is an AI framework designed to enhance the quality of responses generated by large language models (LLMs) \cite{lewis2020retrieval}.
RAG allows LLMs to build on a specialized body of knowledge to answer questions more accurately. It’s like giving the model an open-book exam, where it can browse through content in a book, as opposed to trying to remember facts from memory. In our case, this knowledge base would be specific to Android and could include any or all of the following:
\begin{itemize}
    \item Additional code that needs to be fetched from a large codebase
    \item Static / Code analysis results that could prove useful to comment about a specific issue
    \item Documentation and additional information that allows the LLM to garner a better understanding of the system and task at hand
\end{itemize}

We initially focus our efforts on prompt engineering to leverage the LLM’s knowledge for tracing Android vulnerabilities. We attempt to compare the detection capability of the LLM with other strategies on the same set of applications, thereby providing insights on which approach can uncover more vulnerabilities with fewer false positives. We use these insights to continually update and improve the instructions used for prompting over the course of the experiments.

In order to do this we need a benchmark dataset designed to test Android Security Analysis tools. For this study we target smaller apps that replicate individual vulnerability types. However to understand our design objectives we need to structure our pipeline so it can logically be extended to real applications.
In the next section we move on to a case study attempting detection in an app that has multiple seeded vulnerabilities. This involves integrating retrieval mechanisms for obtaining additional context that can significantly enhance the feasibility and performance of such systems. Retrieval mechanism are critical in real-world scenarios when dealing with larger applications due to the limited token window of present-day LLMs. Hence it is also important that we focus on optimising the size of the input to the model while ensuring the LLM has enough context to make a sound decision.

\section{METHODOLOGY} 

\begin{figure*}[h]
\centering
{\includegraphics[width=6.5in]{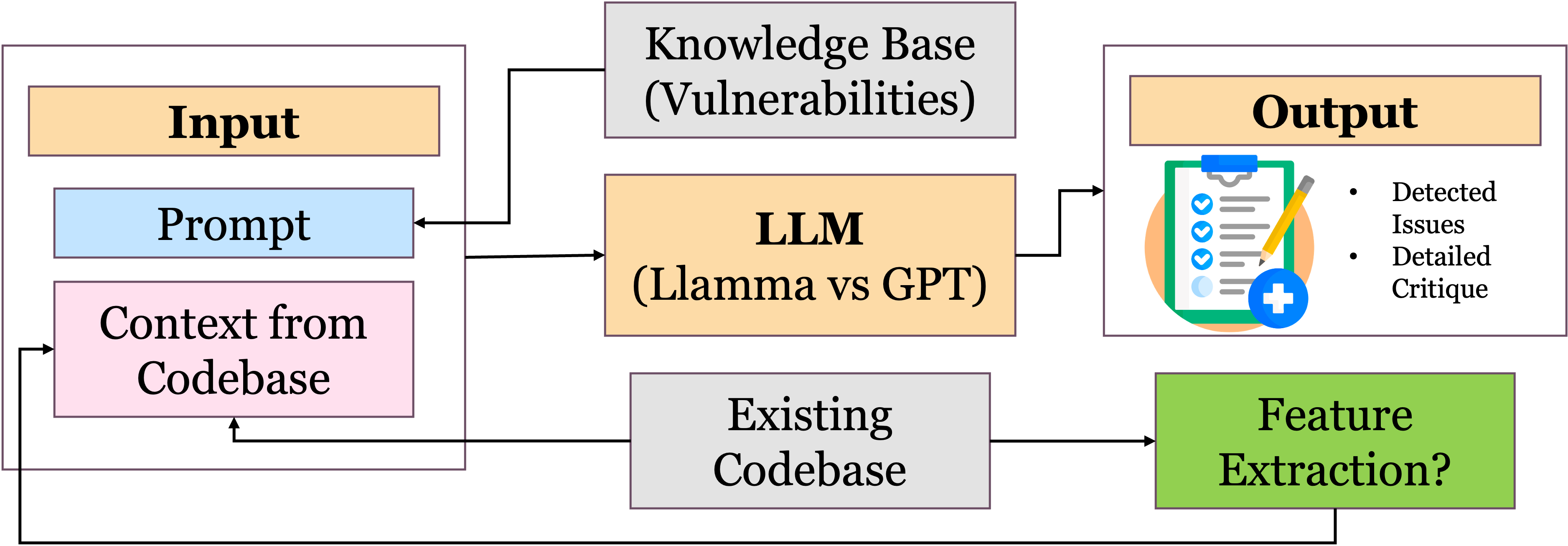}}
\caption{Rough Architecture of Experiments}
\label{fig:arch_diag}
\end{figure*}

In our study, experiments were conducted with GPT-4, using the Ghera benchmark \footnote{\href{https://secure-it-i.bitbucket.io/ghera/index.html}{Ghera Dataset}} as the dataset. We use GPT4 for our experiments as it is one of the most powerful available LLMs as of today. In our limited experiments with GPT3.5 it struggles with following instructions properly. Further we also setup our experiments to work with Open LLMs but as of writing this report open models like Llama 2 perform significantly worse compared to their paid counterparts. Hence we stick with out decision to employ GPT4 despite the higher costs incurred for inferences. Throughout the iterations, we refine the prompts, enhancing the information provided to the model, including specific details about vulnerabilities. In the final experiment, only ``AndroidManifest.xml'' and ``MainActivity.java'' files if they exist in the analyzed project were presented to the model. We also provide a list of additional files present in the project. We also experiment with summarizing the content of files and providing that information to the model as well. We also implemented functionality to allow the model to request the content of specific files if necessary for further vulnerability identification.

Ghera is a benchmark repository that predominantly catalogs vulnerabilities previously identified in Android applications, as established in existing literature. This repository is distinctively organized into benchmarks, each targeting a specific vulnerability. These benchmarks comprise three types of applications: a benign application that possesses the vulnerability, a malicious application designed to exploit this vulnerability, and a secure application that is immune to such exploitation. The benchmarks are practical, including instructions for building and executing the applications, thereby allowing for the empirical verification of the vulnerabilities and their corresponding exploits. Presently, Ghera's collection exclusively consists of 'lean' benchmarks, which are minimalistic apps specifically designed to showcase the vulnerabilities and their exploits, without incorporating additional complex functionalities. Ghera contains 69 different issues as of date that can be categorized into the following types:
\begin{itemize}
\item \textbf{Crypto}: Vulnerabilities in app encryption methods, including block cipher issues and exposed encryption keys.

\item \textbf{ICC}: Risks in app communication components, such as dynamic receivers and intent hijacks.

\item \textbf{Networking}: Security issues in app network communications, including certificate mishandling and MitM attack susceptibility.

\item \textbf{NonAPI}: Vulnerabilities due to outdated libraries and inherited library flaws in apps.

\item \textbf{Permission}: Security risks from unnecessary or weak app permissions leading to potential unauthorized access.

\item \textbf{Storage}: Data storage vulnerabilities in apps, including external/internal storage risks and SQL injection threats.

\item \textbf{System}: System-related vulnerabilities in apps, focusing on privilege escalation and information exposure.

\item \textbf{Web}: Web component vulnerabilities in apps, such as MitM attack risks and code injection in WebViews.

\end{itemize}

During the experimentation process, we found that the Ghera benchmarks contained explicit indications of "Benign" and "Secure" in both the file paths and the content. For our experiments to be valid and to minimize data leakage to the LLM about the ground truth, we replaced these values with "llbezpekymyapp" and "llsezpekymyapp" respectively. However, we include our insights from runs in Experiments 1 and 2 that had the leaked data to see how much it influences the results. Interestingly leakage of the word secure clearly influenced the outcomes and highlighted the need for clearly sanitizing the data from semantic names that can mislead the LLM's reasoning. Do note that folder names still indicate the kind of vulnerability to some extent but this shouldn't be a cause of concern in any of the subsequent experiments. Our analysis of these experiments and their outcomes will be described in the following section. 

\section{RESULTS}
In this section, we discuss the various experiments conducted, the thought processes behind them, and the insights obtained.

\subsection{Experiment 1: Basic Prompting GPT4}

For this experiment, we explore basic prompting without providing any information about the vulnerability. We wish to figure out what kind of vulnerabilities if any GPT could detect without any explanation about the issue. Results are shown in Table \ref{tab:exp1}.

\begin{table}[h]
    \centering
\begin{tabular}{ c||c c||c c }
&  \multicolumn{2}{l||}{\centering Leaked Data} & \multicolumn{2}{l}{\centering Cleaned Data} \\
\hline
  & Insecure & Secure & Insecure & Secure \\ 
 Flagged insecure & 54 & 21 & 58 & 56   \\  
 Flagged secure & 5 & 38 & 1 & 3  \\
 Undecidable & 1 & 0 & 1 & 0 \\
 Overall & 60 & 59 & 60 & 59
\end{tabular}
\caption{Results of Experiment 1}
    \label{tab:exp1}
\end{table}

It is interesting to note that even without any details about the vulnerability GPT4 could flag apps as insecure. However, it seems that the model just has a tendency toward marking apps as insecure as can be seen with secure apps in the cleaned dataset. We note that even though 58 out of 60 insecure are correctly identified with relevant snippets being highlighted 56 of the 59 secure apps are misclassified.

\subsection{Experiment 2: Providing a summary of the vulnerability}
In this experiment, we check if the model can verify the presence of specific vulnerabilities given a very brief summary. We hypothesize that specifying limited details about issues we are looking for would lead to context activation \footnote{\href{https://www.lakera.ai/blog/what-is-in-context-learning}{In-Context Learning}} and enable the LLM to use its existing knowledge towards analysis. Therefore, we extended experiment 1 by providing short information about a vulnerability and the obtained results are shown in Table \ref{tab:exp2}. This process involves creating brief textual descriptions that encapsulate how each vulnerability functions or why an app is susceptible to it. These summaries are crucial as they offer an initial understanding without the need for detailed, app-specific descriptions, which may not be readily available in real-world scenarios. Such summaries might be akin to Common Weakness Enumeration (CWE) descriptions, providing essential insights at a glance.

\begin{table}[h]
    \centering
\begin{tabular}{ c||c c||c c }
&  \multicolumn{2}{l||}{\centering Leaked Data} & \multicolumn{2}{l}{\centering Cleaned Data} \\
\hline
  & Insecure & Secure & Insecure & Secure \\ 
 Flagged insecure & 53 & 6 & 55 & 34   \\  
 Flagged secure & 6 & 52 & 3 & 24  \\
 Undecidable & 1 & 1 & 2 & 1 \\
 Overall & 60 & 59 & 60 & 59
\end{tabular}
\caption{Results of Experiment 2}
    \label{tab:exp2}
\end{table}

\begin{figure*}
    \centering
    \includegraphics[width=\textwidth]{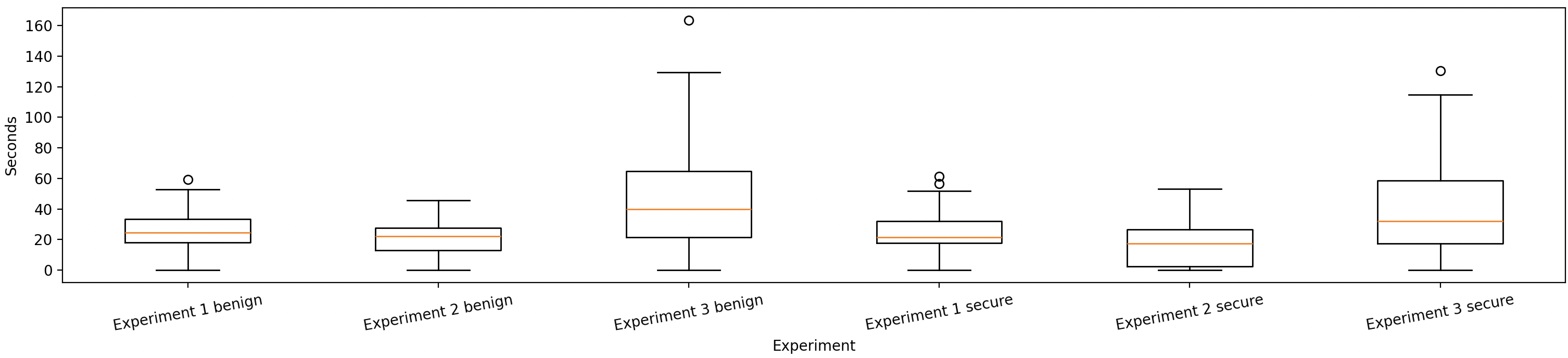}
    \caption{Time consumed for each inference}
    \label{fig:time}
\end{figure*}

We see that with just a short summary of the vulnerability we are able to improve on the number of Secure apps flagged as insecure by a significant margin. Another thing that can be noted from Figure \ref{fig:time} is that Experiment 2 actually reduces the time consumed by the model to return an inference both on secure and benign apps. We notes than we have a TP rate of 66.38\% in this case as compared to 51.26\% in the previous experiment (we consider undecidable as a misclassification).

\subsection{Experiment 3: Requesting files as and when required}
For previous experiments we were sending all the needed files, we cannot use such techniques for real-world applications. Hence, we modified the prompt, so the model asks to provide more information if needed. Results are shown in Table \ref{tab:exp3}. We initially focus on the AndroidManifest.xml and MainActivity.java files, recognizing that while vulnerabilities often reside deeper in the code, these files can provide vital starting points for analysis. From here, the model can request additional files as needed, thereby initiating the Retrieval Augmented Generation process in a structured manner.

Our approach includes providing a summary of all files in the application, with the exception of the Manifest and the MainActivity file. The summary is generated by the LLM itself in a preprocessing step. The LLM is then enabled to request the content of specific files based on these summaries. This selective process is particularly useful in identifying vulnerabilities that might be indicated by exported components or specific implementations in the code. This methodology is a form of Retrieval Augmented Generation though at a very coarse granularity, aiming to probe its impact on the efficiency and accuracy of vulnerability detection.

The undecidable case in this experiment requires the build.gradle file and the "libs" folder and even though that is correctly requested we currently only include code within "app/src/main" and address this in the final architecture. We noticed that the model tends to ask for all files when implemented in a simple fashion so we ended up creating an alternate chain that summarizes all the files and includes this summary with the list of files. This actually had the opposite effect with the model never requesting any file in most cases. We justify the lower TP rates of 60.5\% due to hallucinations made by the model based on the summaries. Another thing to note is that even though the impact is not significant, the generated reports lack clarity due to not having access to the exact implementation details. However there are clearly efficiency gains to an approach like this though while sacrificing on the quality of the report generated.

\begin{table}[h]
    \centering
\begin{tabular}{ c||c c }
& \multicolumn{2}{l}{\centering Cleaned Data} \\
\hline
  & Insecure & Secure \\ 
 Flagged insecure & 57 & 43    \\  
 Flagged secure & 2 & 15   \\
 Undecidable & 1 & 1   \\
 Overall & 60 & 59 
\end{tabular}
\caption{Results of Experiment 3}
    \label{tab:exp3}
\end{table}

\tipbox{\textbf{Key Takeaways:} From our experiments we see that given sufficient context GPT4 can successfully identify vulnerabilities. Summarizing clearly saves significantly on the costs, since the summaries can be used across scanners and is a one-time cost. Attempts at reducing the amount of information still need work and we need a critique process to deal with the models tendency to mark applications as insecure.}

\section{LLB ANALYZER}

\begin{figure}
    \centering
    \includegraphics[width=\linewidth]{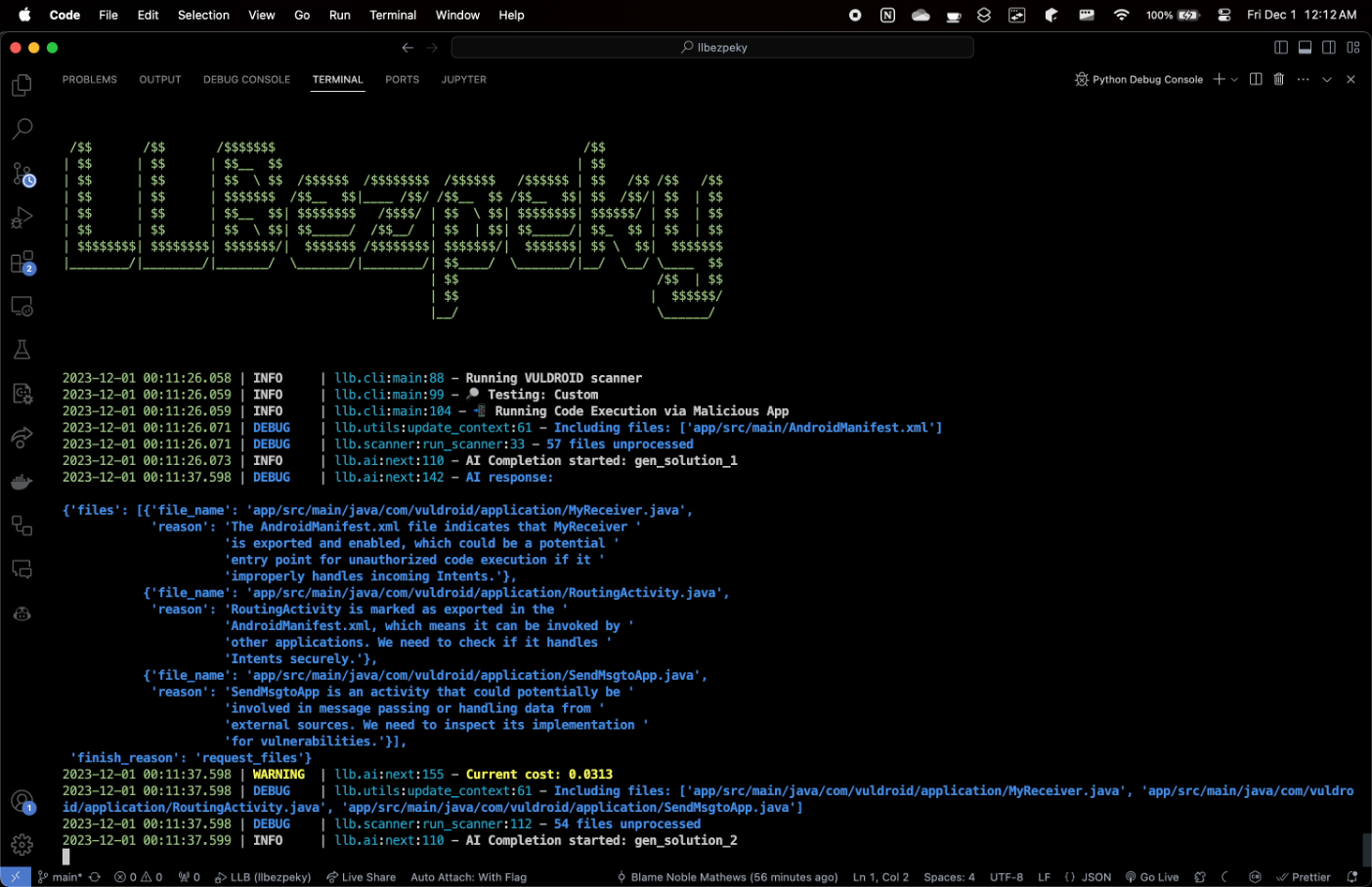}
    \caption{LLB Python package running through CLI}
    \label{fig:llb-terminal}
\end{figure}

We use the insights from the experiments to put together a python package called ``LLB'', which can be invoked to run the the pipeline on any target android application. The package is designed to facilitate the scanning of Android projects for security vulnerabilities. It employs a Command Line Interface (CLI) as shown in Figure \ref{fig:llb-terminal}, leveraging the Typer framework to provide an intuitive user experience. This package integrates distinct scanning mechanisms, offering flexibility and comprehensiveness in the vulnerability assessment process.

\begin{figure}
    \centering
    \includegraphics[width=\linewidth]{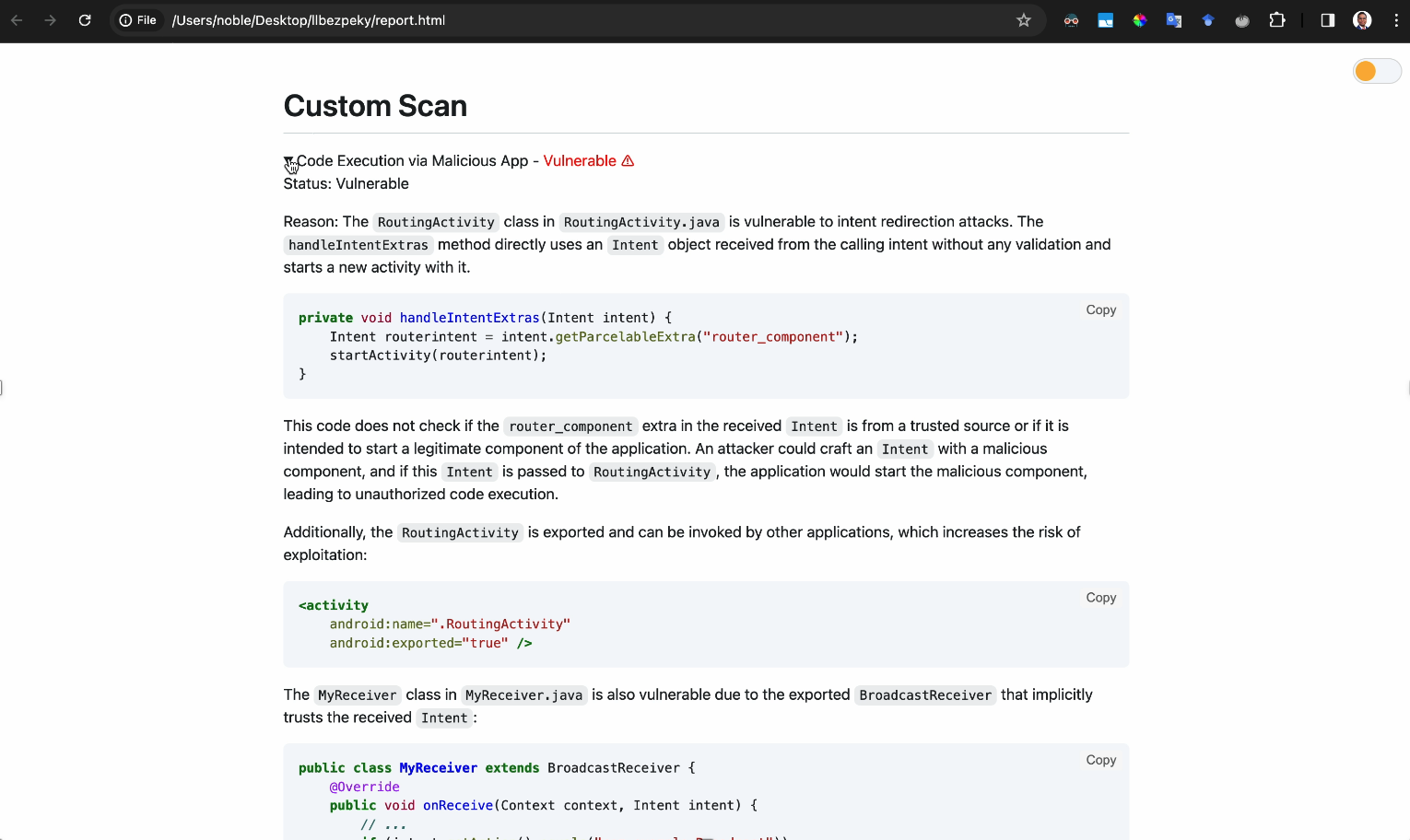}
    \caption{Example from case study report generated by LLB}
    \label{fig:llb-report}
\end{figure}

The core functionality of "llb" is encapsulated in the scan command. This command allows users to specify the target Android application directory for analysis. In alignment with the package's focus on adaptability, it incorporates multiple scanner options, namely 'GHERA' and 'VULDROID', along with an 'all' option to execute all available scanners. These scanners are tailored to identify different categories of vulnerabilities, ensuring a thorough examination of the target application. The scan command also supports an output directory specification, where the results of the scan are systematically compiled and stored. This feature is critical for maintaining a record of the vulnerabilities identified and serves as a reference for further analysis or remediation efforts. An example of a part of the report generated by LLB is shown in Figure \ref{fig:llb-report}.


In addition to its scanning capabilities, "llb" includes the expert command. This command is particularly useful for post-scan analysis, allowing users to append expert comments or insights to the generated reports. This feature underscores the package's utility in collaborative environments, where multiple stakeholders, including security analysts and developers, might interact with the scan results.

\section{Case study}
\subsection{Flagged Vulnerability: Deep dive}

Let us look at the analysis report for case ``WebView NoUserPermission InformationExposure'' in Ghera.
In mobile app security, it's crucial to handle sensitive user data carefully. In this case we have a simple web browser that allows web pages to access the device's location via a GPSTracker API. The API's getLatitude() method can be triggered by JavaScript on a webpage to fetch the user's latitude without asking for the user's consent each time. This creates a vulnerability: a malicious webpage could use this method to secretly track a user's location. To mitigate this risk, apps must not only ask for user permission to access sensitive data initially but also maintain strict control over its access during use, particularly when dealing with potentially untrustworthy web content.

\begin{lstlisting}[float=h, label={code:snippet_1},caption={Information Exposure snippet flagged by LLB as indicated in the Ghera benchmark}, firstnumber=1,language=iJava]
@JavascriptInterface
public double getLatitude() {
    if (location != null) {
        return location.getLatitude();
    } else {
        // return dummy value
        return 42.42;
    }
}
\end{lstlisting}

In this case LLB correctly identifies the concerning snippet shown in Listing \ref{code:snippet_1}. LLB returns the reason as follows: ``The application is vulnerable because it exposes sensitive information (GPS location) to the JavaScript code running in the WebView without explicitly asking the user for permission… The application does request the ACCESS\_FINE\_LOCATION permission, but it does not inform the user that their location will be accessible to any malicious JavaScript code running in the WebView.'' and also provides a suggested fix to remediate the issue.

\subsection{Misclassification: Deep dive}

A common trend which we observed among misclassified samples was that they either were flagged for bad patterns in code which might not necessarily be considered vulnerable by an expert in the context we are analysing. Let us look at a case where the LLM was thrown off. The initial classification of the code as "Vulnerable" due to lack of sanitization checks on the file paths seems to be based on a misunderstanding. In the provided snippet, the file name fileNm is hard-coded to "demo.txt", and there's no indication that it is influenced by external inputs or user-provided data. Therefore, the specific concern of path traversal, where an attacker could manipulate file paths to access unauthorized directories, does not apply here.

\begin{lstlisting}[float=h, label={code:snippet_2},caption={Misclassified example in Ghera}, firstnumber=1,language=iJava]
String fileNm = "demo.txt";
// ...

if (i != null && i.getStringExtra("fileName") != null && !i.getStringExtra("fileName").equals("")
   && i.getStringExtra("fileContent") != null && !i.getStringExtra("fileContent").equals("")
   && i.getStringExtra("fileName").equals(fileNm)) {
    fileContent = i.getStringExtra("fileContent");
} else {
    setResult(RESULT_CANCELED);
    return;
}

// ...

File file = new File(dir, fileNm);
try (FileWriter fileWriter = new FileWriter(file)) {
    fileWriter.append(fileContent);
} catch (IOException e) {
    Toast.makeText(getApplicationContext(), "IOError while writing", Toast.LENGTH_SHORT).show();
    setResult(RESULT_CANCELED);
    e.printStackTrace();
}
// ...
\end{lstlisting}

It's likely that the Language Model (LLM) that made the initial classification may have been "thrown off" due to the common pattern of vulnerabilities associated with file handling operations, especially when they involve input from intents or external sources. When it sees code that interacts with file systems, there may be an overcautious approach to flag potential security risks, such as path traversal.

The expert follow-up correctly clarifies that, in this specific case, there is no direct vulnerability present in the snippet due to the fixed file name. Using the expert comment we are able to set the analyzer on the right track even in other cases and give a more relevant suggestion if required.

\subsection{Vuldroid}

In order to analyze a full application with multiple known vulnerabilities we chose to conduct a case study on Vuldroid. Vuldroid is a vulnerable Android application, which contains only security issues. The app consists of the following vulnerabilities:
\begin{itemize}
\item \textbf{Steal Password MagicLoginLinks (V1)}. This vulnerability allows a malicious app to intercept password reset tokens or magic login links. This is possible because the app doesn't properly restrict which activities can handle deep links, allowing an external app to capture these sensitive tokens.
\item \textbf{Webview Xss via Exported Activity (V2)}. Here, the vulnerability lies in an exported activity that loads web content. Since the activity doesn’t validate the URLs it loads, a malicious app can pass a script (like a JavaScript prompt) through an intent, leading to a cross-site scripting (XSS) attack.
\item \textbf{Webview Xss via DeepLink (V3)}. Similar to the previous one, this exploit involves XSS attacks but through deep links. The app fails to validate deep link URLs, allowing the injection of malicious scripts via query parameters in these URLs.
\item \textbf{Stealing Files via Webview (V4)}. This issue arises because the app’s webview settings allow access to local file URLs (file:///). By crafting a specific URL, an attacker can access and transmit local files to a remote server.
\item \textbf{Stealing Files via Fileprovider (V5)}. The FileProvider is misconfigured to expose all paths, and combined with an exported activity, it allows other apps to access and steal files.
\item \textbf{Intent Sniffing Between Two Applications (V6)}. This vulnerability occurs when two apps communicate using intents without proper security checks. A malicious app can intercept these intents and access the transmitted data.
\item \textbf{Reading User Email via Broadcasts (V7)}. Due to an exported broadcast receiver in the app, a malicious app can trigger this receiver and access the user’s email information.
\item \textbf{Command Execution via Malicious App (V8)}. This vulnerability allows a malicious app to execute unauthorized commands or operations within the vulnerable app. The details of this exploit weren’t fully described in your summary.
\end{itemize}

The results of the case study are described in Table\ref{tab:casestudy}. The meaning of the symbols:
\begin{itemize}
\item \texttt{C} - vulnerability was detected
\item \texttt{W} - the provided information wasn't enough for the LLM to make the decision.
\item \texttt{X} - vulnerability wasn't detected
\end{itemize}

\begin{table}[h]
    \centering
\begin{tabular}{ c|c|c|c|c|c|c|c }
 {\centering V1} & {\centering V2} &  {\centering V3} & {\centering V4} &  {\centering V5} & {\centering V6} &  {\centering V7} & {\centering V8} \\
\hline
  \texttt{W} & \texttt{C} & \texttt{C} & \texttt{C} & \texttt{W} & \texttt{C} & \texttt{C} & \texttt{C}
\end{tabular}
\caption{Results of case study}
    \label{tab:casestudy}
\end{table} 

We note that running LLB on the Vuldroid source code we correctly identify 6 of the 8 seeded vulnerabilities. The 2 which couldn't be classified were because the analyzer could not find the relevant snippets for analyzing by requesting files. For the remaining 6 cases the correct snippet is identified and a valid fix is also suggested in most cases. Thus the LLB report doesn't just tag vulnerable applications but actually walks you through the reasoning involved and how to fix the flagged issue.

\section{DICUSSION AND FUTURE WORK}

The field of leveraging Large Language Models (LLMs) to enhance Software Engineering tools and improve our understanding of large projects is a very active research area. There are numerous parameters and processes that can be optimized based on the specific functionality desired. We explore a basic prompting strategies for LLB however a more structured pipeline with multiple agents could significantly improve the performance of LLB as an analyzer. We also wish to highlight that each vulnerability we scan for is considered as a single scanner and there is clear value to exploring making this process less resource intensive by sharing information between scanners and choosing which ones to run based on the type of codebase and artifacts present within it. 

We also wish to compare the results with those obtained from existing approaches in an empirical study. 
While our current framework does not integrate static analysis, we acknowledge its potential value and are considering its incorporation in future iterations. 
Overarching the entire methodology is a constant focus on improving software vulnerability identification, reducing false positive rates, and streamlining the process of enhancing Android app security. We believe our study serves as a basis to attempts to merge state-of-the-art language models with static analysis, potentially establishing a more reliable, accurate, and efficient android vulnerability detection approach. ``What existing tools and code level analysis results can be supplied to the LLM to take a more informed decision?'' remains an open question.

The results of the LLB case study on vuldroid highlights a shortcoming in the current implementation. This can be easily handled by indexing the code into a vector database and allowing vector search and retrieval to identify relevant files rather than building outwards from a set of files. Further summaries were not incorporated in the LLB package due to their tendency to induce hallucinations in the report. We plan to address this by increasing our granularity to a file level and incorporating critique mechanisms like have been employed in recent works in other domains that attempt to leverage LLMs.

\section{THREATS TO VALIDITY}

Prompt engineering, while a powerful tool to guide LLMs, is also subject to limitations. The effectiveness of prompt engineering is heavily reliant on the skill and experience of the user. Poorly designed prompts can lead to suboptimal results, as the model's responses are only as good as the questions posed. There's also the risk of introducing bias through prompts, which can skew the model's focus and potentially overlook certain types of vulnerabilities.

A point which we highlighted earlier is about leaking semantic information about the class of the problem due to artifacts in the code. We replace keywords to clean our data but there could be implicit data that leaks this information to the LLM which we might not have accounted for in out analysis. Further capability of LLMs is very diverse with some performing drastically different compared to others. While we set seeds to ensure replicability of our results these can vary drastically over time.

Another concern is the dynamic nature of both Android platform and cybersecurity threats. As Android continuously evolves, new types of vulnerabilities emerge, which may not be immediately recognized by an LLM trained on outdated data. Similarly, cyber threats are constantly evolving, with attackers devising new methods to exploit systems. This requires continuous updates and retraining of the LLM, which can be resource-intensive.

\section{CONCLUSION}
In our research, we explore the utilization of Large Language Models (LLMs) for detecting Android vulnerabilities, We successfully demonstrate the power of LLMs for Android Vulnerability detection and remediation. Our experiments using Prompt Engineering on the Ghera Vulnerability Dataset show promising results and bring up new and interesting directions which can be explored towards improving the efficacy of such systems. Further we utilize the results and insights from our experiments to create a highly configurable python package that allows easy modification of the LLM being used as the reasoning engine and also supports extension to multi-agent architectures. In terms of the questions we set out to answer it is clear that LLMs are incredibly powerful tools that can revolutionize Software Engineering tools as we know them, but it also clear that they do not work magic out of the box and clearly require work in terms of drafting and structuring a better analysis pipeline architecture and optimising the context available to the LLM.

\section{ACKNOWLEDGEMENTS}
This project report is a part of the course CS858 at the University of Waterloo. I express my sincere gratitude to my course instructor, Yousra Aafer, for her invaluable guidance and support throughout this project. Her expertise and insights have been fundamental in shaping the research and its outcomes.

I would also like to extend my thanks to my supervisors, Meiyappan Nagappan and Shane McIntosh. Their input and feedback have been instrumental in refining the research methodologies and enhancing the overall quality of this work. Their perspectives and suggestions have greatly contributed to the depth and rigor of the research.

I am thankful for the collaborative environment provided by the University of Waterloo, which has been conducive to academic exploration and innovation. The resources and support offered by the university have played a crucial role in the successful completion of this project.

Lastly, I appreciate the efforts of all those who have directly or indirectly contributed to this research, including my peers for their constructive criticisms and the university staff for their administrative support.

\bibliographystyle{ACM-Reference-Format}
\bibliography{IEEEabrv,proposal}

\end{document}